# The tear turnover and tear clearance tests – a review

Izabela K. Garaszczuk[1], MSc, Robert Montes Mico[1], PhD, D. Robert Iskander[2], PhD, DSc, Alejandro Cerviño Expósito[1], PhD

**Affiliations:**

1. Department of Optics, Optometry and Vision Sciences, University of Valencia, Dr. Moliner, 50, 46100 Burjassot, Valencia, Spain;
2. Department of Biomedical Engineering, Wroclaw University of Science and Technology, Wybrzeze Wyspianskiego 27, 50-370 Wroclaw, Poland

**Corresponding author:**

Alejandro Cerviño Expósito, PhD
Department of Optics, Optometry and Vision Sciences,
University of Valencia,
Dr. Moliner, 50, 46100 Burjassot, Valencia, Spain,
e-mail: Alejandro.cervino@uv.es;

**Correspondence:**

- Izabela K. Garaszczuk, MSc
  e-mail: Izabela.garaszczuk@uv.es
- Robert Montes Mico, PhD
  e-mail: Robert.montes@uv.es
- Robert Iskander, PhD, DSc
  email: Robert.iskander@pwr.edu.pl

# ABSTRACT

**Introduction.** The aim is to provide a summary of methods available for the assessment of tear turnover and tear clearance rates. The review defines tear clearance and tear turnover and describes their implication for ocular surface health. Additionally, it describes main types of techniques for measuring tear turnover, including fluorescein tear clearance tests, techniques utilizing electromagnetic spectrum and tracer molecule and novel experimental techniques utilizing optical coherence tomography and fluorescein profilometry.

**Areas covered.** Internet databases (PubMed, Science Direct, Google Scholar) and most frequently cited references were used as a principal resource of information on tear turnover rate and tear clearance rate, presenting methodologies and equipment, as well as their definition and implications for the anterior eye surface health and function. Keywords used for data-search were as follows: tear turnover, tear clearance, fluorescein clearance, scintigraphy, fluorophotometry, tear flow, drainage, tear meniscus dynamics, Krehbiel flow and lacrimal functional unit.

**Expert Commentary.** After decades, the topic of tear turnover assessment is reintroduced. Recently, new techniques have been developed to propose less invasive, less time consuming and simpler methodologies for the assessment of tear dynamics that have the potential to be utilized in a clinical practice.

**Funding details.** This work was supported by the Horizon 2020 Research and Innovation Program under the Marie Sklodowska-Curie Grant 642760.

**Disclosure statement.** The authors report no conflict of interest

## 1. INTRODUCTION

A well-balanced composition of the tear film and effective tear dynamics play an important role in the ocular surface health and function. Tear dynamics of a healthy eye consists of adequate tear production, unaffected distribution and retention of tears on the ocular surface, tear turnover, elimination through the nasolacrimal duct and tear fluid evaporation and absorption into surrounding tissues [1]. These processes are regulated by the lacrimal functional unit (LFU), which is composed of secreting glands (main and accessory lacrimal glands and Meibomian glands), conjunctival goblet cells, the ocular surface (cornea and conjunctiva) and their interconnecting innervation [2, 3]. To achieve tear homeostasis the LFU ensures the delivery of supportive and protective substances, formation of tear film and clearance of pro-inflammatory or toxic factors through the nasolacrimal drainage system [2, 3, 4]. Blinking generates a pumping effect on the nasolacrimal drainage system to draw tears into the lacrimal sac and plays a key role in tear dynamics by spreading, mixing, and distributing the tears and clearing cellular and other debris [4, 5, 6]. Spontaneous blinks are thought to be modified by reflex inputs from the ocular surface and inputs from the higher centers. The blink cycle consists of the blink itself and the blink interval, during which evaporative water loss occurs. Rapid flow of tears into the canaliculi can be observed during the first few seconds following the blink [7]. The lacrimal drainage capacity increases with increasing blink rate and aging [8]. When the excess fluid is present in the conjunctival sac, the expansion of the lacrimal sac serves as to draw in the excess fluid for many seconds following the blink, resulting in a steady flow towards the lacrimal punctum, called the Krehbiel flow [4, 9]. Since, the canaliculi can normally transport more fluid in one blink that is attributed to normal, non-reflex tear secretion, the drainage system is operating nearly empty most of the time in normal conditions. Blink rate drops after topical anesthesia [10], and following LASIK surgery [11].

Tear turnover is described as a global measure of the integrity of the lacrimal functional unit and tear exchange on the ocular surface [1, 3, 12]. Tear turnover rate (TTR), a temporal measure of tear turnover is proportional to the sum of the effects of tear secretion by the glands (denoted by S), fluid transudation through the conjunctiva (C), tear drainage through nasolacrimal duct (D), evaporation (E) and conjunctival (PC) and corneal (PK) water permeability:

$$TTR \sim (S + C) - (D + E + PC + PK),$$

where ~ denotes the proportionality and D > E > PC > PK in normal circumstances [3, 12].

Tear clearance rate (TCR) can also be defined as a temporal measure of tear exchange on the ocular surface. However this term usually refers to the speed of disappearance of exogenously added dye or other topical substance from the ocular surface measured by other means than fluorophotometry. For example we refer to TCR in cases such as fluorescein clearance test, colorimetric techniques and optical coherence tomography.

TTR was shown to be an indirect measure of dry eye associated ocular surface irritation regardless of normal or reduced aqueous tear production [13, 14, 15, 16] and found to be reduced in symptomatic dry eye subjects [1, 12, 17, 18, 19]. The value of TTR allows to distinguish between aqueous deficient and evaporative dry eye [20]. TTR correlates with the severity of ocular epithelial disease assessed with corneal fluorescein staining [15, 16, 21] rather than with reduced aqueous tear production, often assessed with the Schirmer test [21]. Delay of tear clearance is also associated with Meibomian gland dysfunction [15, 16] and decreased ocular surface sensitivity [3, 14, 15, 16, 22, 23, 24, 25]. In a mouse model of dry eye, reduced tear clearance showed greater correlation with the severity of ocular surface disease than reduced tear production [25]. Also, age and factors associated with age, like conjunctivochalasis, lid laxity, functional obstruction to tear flow, blink abnormalities contribute to development of delayed tear turnover [14, 15, 26, 27, 28]. On the contrary, some studies showed no relation with aging [29, 30, 31]. Reduced tear clearance promotes ocular surface inflammation [14], as it leads to accumulation of cytokine interleukin-1α (IL-1α) and the activity of matrix metalloproteinase (MMP-9) and gelatinase B in tear film [14, 15, 16, 22, 23, 32], and was proved to improve with topical methylprednisolone, together with decrease in symptoms of ocular irritation, conjunctival redness and surface epithelial disease [14, 15]. However, the exact mechanism by which reduced clearance leads to corneal epithelial disease and ocular irritation has not been established. It is suggested that the increase in MMP-9 activity may lead to corneal epithelial disease. Tear clearance is also reduced in subjects with contact lens associated papillary conjunctivitis [33]. Delayed tear clearance may be the best measure for identifying patients with tear film disorders, who may respond to anti-inflammatory therapy [3]. Delay in tear clearance can lead to prolonged exposure to topical medications and their preservatives (e.g., benzalkonium chloride) on the ocular surface compared with normal clearance subjects, thus affected individuals have higher chance to develop ocular surface medication toxicity [12, 14, 15, 34]. Decreased tear clearance is also associated with untreated [35] and timolol-treated open-angle glaucoma [36].

2. **Techniques used in tear turnover and tear clearance assessment**

In general, tear turnover rate or tear clearance rate are estimated based on direct or indirect observation of the decay of the dye placed in the tear fluid over time with the production and elimination of tears. It can be assessed in vivo or in vitro. These techniques originated from the one used by Nover and Jaeger in 1952 [37]. The most popular approach is based on following the elution of tracer molecule from the tear film. [1, 12, 18, 30, 31, 38, 39, 40, 41, 42, 43, 44, 45, 46] This family of methods include, the most commonly used – fluorophotometry and gamma scintigraphy [1, 5, 47, 48, 49, 50, 51]. In case of fluorophotometry, the molecule is excited with external radiation whereas in lacrimal gamma scintigraphy the emission of radiation is caused by the natural disintegration of an atom.

## 2.1. Fluorophotometry

Fluorophotometry is considered the gold standard in tear turnover and tear flow assessment [50, 52]. After the first use of fluorophotometry in 1954 to study aqueous humor dynamics, the device was gradually modified and improved, until the first commercially available fluorophotometer (Palo Alto-based medical division of Coherent, Inc.) was introduced in 1980. It evolved to be a laboratory device used to quantify the fluorescein concentration in the eye. For tear turnover estimation, the modified slit lamp fluorophotometers [28, 40, 44, 45, 52, 53, 54] were in use before commercially available tools were developed. Some attempts to standardize the procedure were made [30] together with the development of a custom-written software for data processing ("ANT_SEGMENT tear" by Leiden Centre for the European Concerted Action on Ocular Fluorometry [30]).

A schematic diagram of a basic fluorophotometer has been displayed in figure 1. This device utilizes a beam of light passing through and excitation filter (430-490 nm) to excite the transparent layers in the eye. The light emitted from ocular tissues passes through a fluorescence emission filter (510-630 nm). The amount of fluorescence is measured by a photomultiplier in the area called the focal diamond, where the excitation and emission beam reach, passing through the same lens system. The anterior segment attachment has to be used to reach the area of interest, as the focal diamond is then moved along the optical axis of the eye, by moving the lens system with a stepped motor. Disodium fluorescein is the most commonly used fluorophore in TTR assessment. It is a hydrophilic molecule, with low topical toxicity and with excitation wavelengths of 475–490 nm and emission wavelengths of 510–520 nm [12].

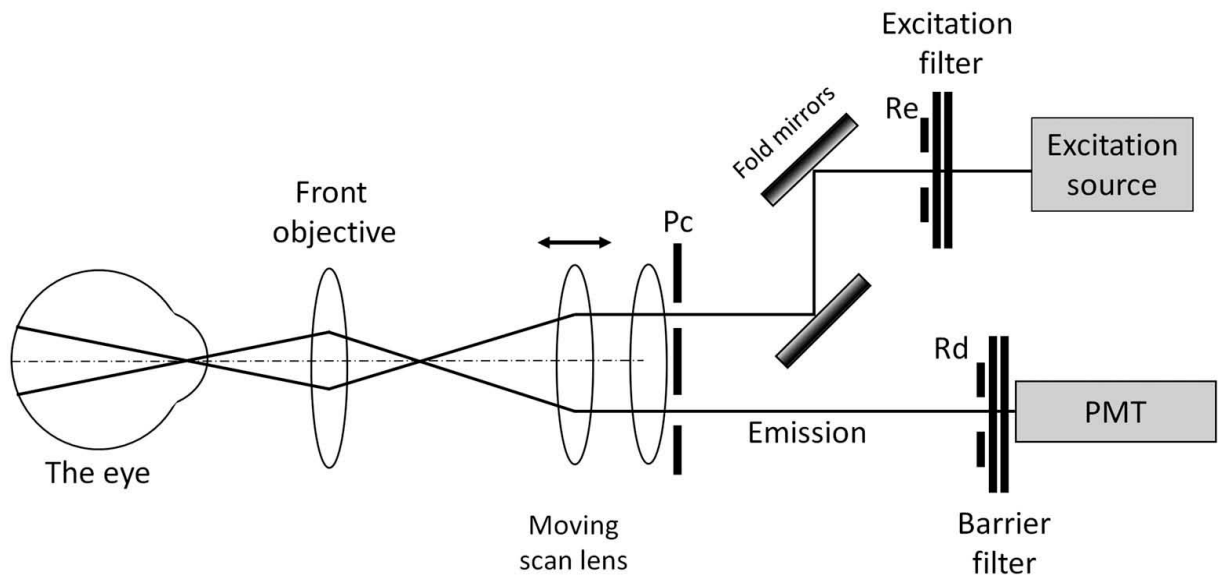

Figure 1. Schematic diagram of a basic fluorophotometer. Re, Rd – slit apertures, Pc – excitation and detection pupils, PMT – photomultiplier tube.

Standardized procedure [30] of tear turnover assessment (described in 1995) lasts up to 30 minutes [24]. Fluorophotometric scans are performed every two minutes with a commercially available fluorophotometer designed for in vivo analysis (Fluorotron Master, Coherent Radiation Inc. Mountain View, CA, USA) following an instillation of 1 μl of 2 % sodium fluorescein into the subject's lower conjunctival sac with a micropipette.

Generally in TTR studies the instilled volume of fluorescein varies among studies (from 1 up to 5 μl), as well as fluorescein concentration (up to 10%) and the sampling rate and duration of the procedure (10 up to 30 minutes). Continuous fluorophotometric measurements have also been performed [38]. In vivo fluorophotometric measurements of fluorescence decay can be performed on the tear meniscus or the precorneal layer of the tear film.

### 2.1.1. Biphasic characteristics of tear turnover

In most of the kinetic studies of tear turnover with fluorophotometry the biphasic characteristics of tear clearance is observed [1, 12, 30, 36, 38, 52], with faster phase occurring just after fluorescein instillation, which seems to be caused by reflex lacrimation produced when fluorescein drop is added to the tear film [1, 12, 14, 26, 51, 52]. The first phase lasts approximately 5 minutes after instillation and varies from subject to subject. It was found to be correlated with subjective signs of irritation at the time of instillation [52], can be suppressed by anesthetics [14, 26] and highly depends on blinking frequency [26, 38, 51, 55]. Studies also show that elderly subjects, especially women, tend to lose the initial phase of tear clearance and have generally lower tear clearance rates [28, 52, 56]. The second, slower phase presumably represents the tear turnover under basal conditions of secretion and is used to calculate basal tear turnover rate [12, 30, 39]. Some studies suggest that only the first phase decreases with age, while the physiological tearing stays the same and some studies report no change of TTR with aging [12, 52]. This basal part of the curve is fitted with an appropriate software [30] and the decay in fluorescence is calculated from the log of the curve obtained from the formula below:

$$T_0(t_0) = \frac{100\,[C_t(t_o) - C_t(t_o+1)]}{C_t(t_o)} \;[\%/min],$$

where $C_t(t)$ represents the fluorescein concentration in the tear film at time $t$[min], while $t_0$ represents any given moment after instillation. Assuming a monophasic decay of fluorescence after initial 5 minutes with a decay time constant $\beta(min^{-1})$:

$$C_t(t) = C_t(0)e^{\beta t} \;[ng/ml]$$

we can calculate

$$TTR(t_0) = 100\left(1 - e^{\beta t}\right)\;[\%/min].$$

The rate of decay of tear film fluorescence is recorded as % per minute. In order to express the turnover value in terms of [μl/min] (often referred to as "flow") it is necessary to assume the tear volume [53, 57] or to measure the volume from the initial dilution of fluorescein, which is calculated by back-extrapolation to time zero of the initial fluorescence decay [1, 30, 36]. Farris et al. reported wide variation in tear film production in dry eye subjects and difference between dry eye and normal subjects in basal tear volumes [58, 59].

Table 1. shows a summary of tear turnover rate values calculated by means of fluorophotometry reported in literature from 1980 (since the first commercially available fluorophotometer was used) until recent times. Most of these studies show that tear turnover rate can be used as a measure for the differential diagnosis of dry eye disease. Large variability of reported results can be due to different methodologies, groups of subjects and diagnostic criteria applied. TTR can have relatively large standard deviations, suggesting large variability amongst subjects. Values reported for dry eye subjects are two to five times lower than values reported for controls [18, 31]. Studies show that TTR may have a potential to distinguish between dry eye subtypes [20]. Meta-analysis conducted by Tomlinson et al. showed a 50% reduction in TTR in aqueous deficient dry eye subjects and 25% reduction in evaporative dry eye subjects with respect to healthy subjects [60]. Studies of Webber et al. [39] showed that tear turnover was found to increase suddenly due to a variety of stimuli, like coughing, sneezing, wind or even psychological factors, like the effort to keep the eyes open [52], as well as variation in blinking rate in some individuals. The rate of spontaneous blinking [61] is adapted to environmental conditions and varies highly even with normal subjects, reflecting individual variation and the influence of environmental and experimental conditions. Blink rate is strongly influenced by mental state, attention, physical activity, eye exposure, relative humidity and temperature of the environment and airflow over the eye and falls during a number of common visual tasks requiring mental concentration. It is increased by low humidity, cold and high wind speeds. Blink rate is found to increase in DED, where it is thought to play a compensatory role in refreshing the tear film more frequently [264,265].

Webber and Jones also suggested a circadian rhythm of TTR with significantly higher values reported in the morning [38, 40, 62, 63].

Table 1. Group average values (mean ± standard deviation) of tear turnover rate measured in vivo with fluorophotometry reported in the literature from year 1980.

| Year | Report | Type (Number of subjects) | Reported TTR [%/min]. Mean ± S.D. |
|---|---|---|---|
| 1980 | Puffer et al. [45] | Normal (51) | 15.4 ± 11.9 |
| 1980 | Jordan, Baum [26] | Normal (18) | 12 ± 3 |
| 1987 | Webber, Jones [40] | Normal (16) | 14.9 ± 5.6 |
| 1988 | Occhipinti et al. [31] | Normal (12) | 27 ± 14 |
| 1992 | Kuppens et al. [36] | Normal (27) | 15.7 ± 5.3 |
|  |  | Timolol- treated glaucoma (24) | 10.1 ± 3.2 |
|  |  | Untreated glaucoma (13) | 12.3 ± 4.1 |
| 1992 | Kok et al. [64] | Normal (25) | 15.2 ± 4.9 |

| 1992 | Goebbels et al. [18] | Normal (20) | 22.2 ± 0.9 |
|---|---|---|---|
| | | DED (20) | 6 ± 6 |
| 1995 | Van Best et al. [30] | Normal (48) | 16.4 ± 4.4 |
| 1995 | Nelson [12] | Normal (13) | 19.25 ± 7.70 |
| | | DED (13) | 11.23 ± 5.38 |
| 1995 | Xu and Tsubota [39] | Normal (34) | 10.7 ± 8.0 |
| 1996 | Sahlin, Chen [29] | Normal (43) | 10.9 ± 3.1 |
| 1996 | Mathers et al. [19] | Normal (72) | 7 ± 4 |
| | | DED (37) | 5 ± 3 |
| | | MGD (109) | 7 ± 6 |
| 2001 | Tomlinson et al. [65] | Normal (20) | 21.4 ± 11.1 |
| 2001 | Tomlinson et al [66] | Normal (9) | 16.6 ± 6.7 |
| 2001 | Sorbara et al. [17] | Normal (10) | 11.85 ± 3.31 |
| | | Symptomatic DED (10) | 4.89 ± 2.75 |
| 2001 | Pearce et al. [42] | Normal (56) | Old method: 12.65 ± 6.64 |
| | | Normal (49) | New method: 17.78 ± 6.34 |
| 2002 | Keijser et al. [67] | Normal (16) | 14.3 ± 6.5 |
| 2003 | McCulley et al. [68] | Normal (22) | 16.3 ± 7.3 |
| | | DED (35) | 13.5 ± 9.3 |
| 2004 | Sorbara et al. [17] | Symptomatic DED (10) | 4.89 ± 2.74 |
| | | Asymptomatic DED (10) | 11.85 ± 3.31 |
| 2005 | Khanal, Tomlinson [1] | DED (8) | 8 ± 3 |
| | | MGD (6) | 11 ± 6 |
| 2008 | McCann et al. [69] | Normal (15) | 20.60 ± 9.3 |
| 2009 | Mochizuki et al. [70] | Normal (12) | |
| | | TTR of the aqueous layer | 10.3 ± 3.7 |
| | | TTR of the lipid layer | 0.93 ± 0.36 |
| 2009 | Tomlinson et al. [60] | Normal (187) | 16.19 ± 5.1 |
| | Meta-analysis | DED (197) | 9.26 ± 5.08 |
| | | Aqueous deficient DED (83) | 7.71 ± 1.02 |
| | | Evaporative DED (94) | 11.95 ± 4.25 |
| 2010 | Khanal et al [20] | Normal | 15.24 ± 5.69 |
| | | Aqueous deficient DED | 5.74 ± 3.69 |
| | | Evaporative DED | 12.61 ± 7.56 |
| 2012 | Khanal et al [71] | Chronic GVHD (12) | 5.40 ± 3.74 |
| | | SS (12) | 5.98 ± 3.75 |
| | | MGD (12) | 12.61 ± 7.56 |

DED – Dry eye disease, MGG – Meibomian Gland Dysfunction, SS – Sjögren Syndrome, GVHD – graph-versus-host disease;

Studies also show the possibility to assess tear turnover rate of lipid and aqueous layer of the tear film, independently [70, 72], utilizing a free-fatty acid conjugate of fluorescein (the

5-dodecanoylaminofluorescein, DAF). Lipid layer has much lower turnover rate than the one of the aqueous component. The limitation of this study is that DAF emulsion has different biophysical properties (thus different TTR) to that of the human lipid layer [72], moreover, a recent study on tear film dynamics applying quantum dots suggested that only the aqueous component of tears exits via the puncta [73].

### 2.1.2. In vitro fluorophotometry

The focal diamond of a basic fluorophotometer is 50 μm wide, 1.9 mm high and around 0.5 mm deep. Its depth approximates corneal and tear film thickness, limiting the spatial resolution of the device, and makes it difficult to determine whether the readings are coming from the tear film or the portion of the cornea. In some individuals, especially dry eye subjects, the precorneal tear film may break-up rapidly exposing the corneal tissue. Additionally, compromised corneal epithelium of these subjects was proven to be excessively permeable to fluorescein, increasing corneal fluorescence [12]. Hence, it is sometimes desirable to perform the assessment of TTR on collected tear samples [15, 16, 23, 24, 63]. Most commonly it is achieved 15 minutes after instilling 5 μl of 2% sodium fluorescein into the inferior conjunctival sac, when fluid is carefully collected from the inferior tear meniscus with a low protein-binding polyester rod or capillary pipette and assessed with a commercially available in-vitro fluorophotometer. [15, 16, 23, 24, 74]. Collected samples volume is determined by comparing the weight of an empty rod to the weight of the rod after sampling [23, 74]. Carefully prepared tear samples can be stored for up to 24 hours without significant change in fluorescence [15]. This technique is however confined to research settings due to an elaborate methodology [1]. The concentration of fluorescein can be described as units of fluorescein per collected volume (units/μl). Studies suggest 274 units/μl to be the threshold between normality and abnormality (80% specificity and 85% sensitivity), while fluorescein concentration of 525 units/μl has 100% specificity for ocular irritation symptoms [15]. Studies based on tear sampling also suggest that reduced aqueous tear production is not the only factor responsible for the delay in tear clearance in symptomatic subjects.

### 2.1.3. Limitations

Irrespective of its popularity, fluorophotometric assessment of TTR still has its limitations. It requires considerable skill to perform measurements, extensive period of time is required to obtain results (from 10 [42] up to 30 minutes [30]). Due to these limitations it has been mostly confined to research settings. The aperture of a photometric microscope is larger than the tear film thickness, which lowers the spatial resolution of the device [38] and its ability to measure the fluorescence coming from a thin tear film layer without including a significant part of the corneal tissue. This leads to errors in TTR estimation due to corneal permeability to sodium fluorescein, especially in subjects with compromised epithelium (subjects treated for glaucoma with timolol and dry eye subjects) [12, 36, 38, 46, 75, 76, 77]. Therefore, a 30 minutes limit for fluorophotometric measurements of TTR is recommended. After 30-40 minutes post instillation the corneal fluorescence starts to influence the results and readings show much lower rate of decrease than is characteristic to tear flow [12, 36, 38]. To address this limitation, Joshi et al. reported a novel technique for sequential measurements of tear turnover and corneal epithelial permeability to fluorescein [75]. They determined that 2 µl of 0.75% fluorescein concentration provides the most reproducible estimates.

Studies suggest that corneal permeability to fluorescein decreases in dry eye subjects undergoing treatment with unpreserved artificial tears, as a result of improvement in corneal epithelial disease [76, 77, 78]. This effect might be counteracted by preservatives, which increases corneal permeability [77, 78, 79]. Also, the inherent auto fluorescence of cornea has to be taken into account and subtracted from the measured corneal and tear film fluorescence when TTR is calculated. Corneal auto fluorescence was shown to increase with age [80].

Other factors that may contribute to errors in TTR are connected to the characteristics of a tracer molecule. Too high concentrations of fluorescein instilled can cause quenching and absorption of light by fluorescein without re-emission [38, 40, 75, 81, 82] or counting saturation of the device electronics [30].

Webber and Jones show significant underestimations of TTR while using 10% fluorescein solutions, when investigating the effect of dye extinction and corneal absorption on the accuracy of TTR measurements with fluorophotometry [38]. Near zero TTRs were observed in subjects when using higher concentrations [45]. Pearce et al. [42] stated that the variations in blink rate and non-uniform distribution of fluorescence intensity over the eye surface during scans may also contribute to errors in TTR calculation [38, 42], together with reflex tearing occurring due to prolonged exposure to excessive facial illumination [42]. Attempts have been made to improve the precision of basal TTR assessment, investigating the minimum time period required for the TTR estimation with an automatic improved fluorophotometric technique, incorporating minimum count of six measurements for a total of 10 minutes to obtain reliable results [42].

Also, the time necessary to take a single measurement may contribute to inaccurate TTR estimations. Scans of the tear film, anterior eye, anterior chamber and crystalline lens last approximately 20 seconds. Research suggests shortening the fluorophotometric data acquisition to 8 seconds [35, 64, 83] to minimize a chance of blink occurring during tear film scanning or to elicit a blink rate to ensure constant tear film thickness [17, 38, 45]. Fluorophotometric measurements have to be made at a consistent and fixed time after the blink. On the other hand, in accordance with the observations made by Doane [84] and Cruz [61] fixed blinking rate would surely affect the quality and the characteristics of blinks and ultimately lead to errors in TTR estimation. Doane showed that a significant portion of the Bell's movement can be observed during a forced blink, while no rotation of the globe occurs during a normal, spontaneous blinking [84]. The voluntary blinks are usually more complete and more uniform in quality than the spontaneous ones [8].

### 2.2. Lacrimal scintigraphy

Lacrimal scintigraphy (or scintillography), is the less commonly used technique for tear turnover estimation by means of electromagnetic spectrum. It was first described by Rossomondo et al. in 1972 as an alternative for radiopaque dye studies of the nasolacrimal system (dacryocystograms) used beforehand [51]. This technique allows quantitative assessment of tear turnover on the eye surface [85] as well as direct observation of tear drainage though the nasal cavity [47, 48, 49].

It involves application of a radioactive tracer, gamma-emitting marker molecule which contain technetium 99, usually in a form of sodium pertechnetate drop (0.013 ml) [47, 48] placed on the marginal strip or on the central part of the cornea [5]. Technetium is used, because of its availability and relatively small dose of radiation, resulting from a short half-life. Usually, the distribution of the tracer is imaged serially (one image per 10 seconds) by a gamma counter camera, for a duration of one minute and then at less frequent intervals, until the entire tracer has drained into the nasal cavity. It is followed in a form of a two-dimensional image representing the distribution of a tracer in the eye and adnexa. The rate of change and the transit time of the tracer through the system provide an estimate of TTR [47, 48, 49, 85, 86]. The radiation coming from different areas can be selected and determined. In scintigraphy, usually the entire region of the exposed ocular and ocular appendages surface is selected as the region of interest, as technetium can be observed from precorneal tear film, the marginal strip, the conjunctival fluid, mucus or gel that collects in the inner angle of the eye, material dried out on the lashes or lids [50] and fluid drainage through the nasolacrimal duct and canaliculi [5]. The progression of radioactive tears through drainage system is followed.

Few different models have been used to analyze the tear flow data based on this imaging technique [49]. Linear compartmental model assumes that the tears are flowing with the same pace through all the different compartments of the nasolacrimal drainage system. Later on it has been shown that variable tear flow is a normal feature of the drainage facility, and does not follow a linear model [48, 49]. A compartmental model was proposed incorporating data from separate components – conjunctival sac, lacrimal sac, the nasolacrimal duct and nasal cavity [49]. This model has been used to qualitatively assess tear flow. A mean tear turnover of 0.45 μl/min was estimated based on the assumption that the slow component of clearance from the conjunctival sac represents the net fractional turnover of that compartment and taking a compartmental volume of 7 μl. Based on an alternative four compartmental model calculations, the tear flow was estimated at 1 up to 8 μl [49]. Craig and Tomlinson were able to measure values of reflex and basal turnover as 3.33±1.95 μl/min and 0.56±0.32 μl/min respectively, using a single compartment model for decay of the radioactive tracer on the conjunctival surface [85].

Lacrimal scintigraphy provides a visual evidence for tear drainage. However, it is an invasive and relatively expensive technique, since it requires radioactive substances to be used, thus it is less likely used in clinical setting compared with fluorophotometry.

### 2.3. Fluorescein tear clearance tests

Fluorescein tear clearance tests are simpler and less time consuming techniques that do not require such considerable skills to be executed as the techniques discussed earlier. Up to date, few techniques were developed which are based on the observation of fluorescein dye clearance [3, 13, 14, 21, 23, 24, 39, 87]. Fluorescein clearance tests are based on an observation of how quickly the dye appears in the nasal cavity by the use of a simple cotton tip [88, 89], by direct visualization of fluorescein drainage through endoscopy [90] or by measuring how slow the dye is retained on the ocular surface by means of visual semi-quantitation [37, 91, 92]. However, it was reported that in more than 20% of normal subjects the dye was not recovered in the nasal cavity [91]. It agrees with the suggestion by Maurice [6] and Doane [4] that the low volumes of tears drawn into the drainage system during normal, non-reflex tearing can be absorbed by mucosal surface of the nasolacrimal duct.

Different tests have been developed to estimate tear clearance in a simplified manner by mean of a Schirmer strip. The strip is used to visualize how quickly the color fades in the tear fluid and how long the dye is retained on the ocular surface [13, 14, 27, 32, 39, 89, 91, 92]. Fluorescein tear clearance tests are mostly based on visual comparison of fluorescein concentration collected on Schirmer strips to photographic standards [13, 14, 16, 27, 32]. Fluorescein clearance tests scores were found to correlate with total corneal fluorescein staining scores [15].

Fluorescein Schirmer test can be performed with or without anesthesia [12, 14, 21, 39, 87]. A subject is asked to keep their eyes open for 5 minutes after instilling a 10 µl drop of 0.5 % fluorescein and 0.4 % oxybuprocaine hydrochloride solution into the conjunctival sac [39, 87]. Then, the Schirmer strip is inserted into both eyes and the subject is asked to keep the eyes closed for another 5 minutes [1]. The length of the wet portion is estimated as for the standard Schirmer test together with the tear clearance rate determined by comparing the staining intensity of the Schirmer strip with a dilution standard graded from 1 to 1/256 [39]. Each grade showed a 12.5% increase in basal TTR.

TCR measured this way was proven to correlate with in vivo fluorophotometric assessment of the basal tear turnover [15, 39].

### 2.3.1. Standardized visual scale

There are two main limitations of the fluorescein Schirmer test. It is difficult to compare color on the strip with a photographic standard of liquid fluorescein and the color intensity on the strip is affected by the length of wetting. Hence, is desirable to compare fluorescein stained meniscus directly with a photographic standard. Standardized visual scale method is a colorimetric comparative method [24, 92].

Fifteen minutes post instillation of 5 µl of 2 % fluorescein, the lateral third portion of the lower tear meniscus is compared with one of the colors on a standardized visual scale, with scores ranging from 0 to 6 (normal to highly delayed tear clearance). A score of 3 corresponds to a fluorophotometric value of 274 units/µl, which was reported as a threshold between normality and abnormality [15]. The score correlates with irritation symptoms, ocular surface sensitivity, Schirmer I test scores, the severity of ocular surface, eyelid and Meibomian glands dysfunction (MGD). Its ability to distinguish between dry eye and MGD subjects can be enhanced by using the Schirmer 1 test score [16].

Although tear clearance tests are simple, inexpensive and less time-consuming than previously described methods, they have apparent limitations. Measurements made with these techniques are not direct, often subjective [92] and do not allow following dynamic changes occurring in the tear film nor provide a quantitative measurement of fluorescein concentration.

### 2.4. Experimental techniques

New studies show that there is still room for improvement in the field of tear turnover assessment and new techniques are still being developed to assess TTR due to the need to develop more clinically applicable, objective and less time-consuming methodologies. These techniques include measuring tear meniscus dynamics with optical coherence tomography [93, 94, 95] and tear fluorescein wash-out rate by means of fluorescein profilometry [96].

#### 2.4.1. Optical coherence tomography-based measurement of tear clearance rate

Zheng et al. proposed a new method for early-phase tear film clearance assessment based on following dynamic changes in tear meniscus morphology by means of anterior segment optical coherence tomography (OCT). This technique was used to study a decrease of tear clearance as a function of age [93] and was compared with the fluorescein clearance test results. In that study 5 µl of 0.9% saline solution was instilled to the eye and changes in tear meniscus morphology (tear meniscus height – TMH and tear meniscus cross-sectional area -TMA) over time were followed [93]. The OCT method of TCR evaluation is less invasive and relatively short (5 minutes). OCT ensures good repeatability and allows following changes of tear meniscus morphology during blinking and after topical instillation [97, 98, 99, 100] and is a promising tool for studying the impact of blinking on tear dynamics [101], which proved to be reduced in dry eye subjects [102]. OCT-based TCR was correlated positively with fluorescein clearance tests scores and negatively correlated with the distance between the lacrimal punctum and Marx line, degree of conjunctivochalasis and a degree of lacrimal punctum protrusion. OCT-based measurements show that the lower tear meniscus height and radius have the highest diagnostic value [103, 104]. The most significant decrease in TMH and TMA occurs 30 seconds after saline solution instillation, which suggests that with OCT one can analyze the first, early phase of tear turnover, which is most probably connected with the phenomenon of Krehbiel flow [7]. OCT-based measurements of TCR were shown to correlate with more clinically applicable tests (tear film break up time and tear film osmolarity) [94]. Custom-written software has been developed to enhance the precision of the tear meniscus morphology measurements with OCT and observe the effect of tear meniscus non-confluence after each blink [105]. Recently, a novel technique of contrast-enhanced OCT-based imaging was used to study the clearance of lipids [106]. Table 2 shows that the TCR estimates assessed with OCT are higher than the TTR values assessed with fluorophotometry. Studies suggest that the thin-film dynamics of the tear film that is perched on the ocular surface cannot be compared with the tear meniscus dynamics. The presence of the "black line" that separates the perched tear film from the meniscus when the fluorescein is applied would suggest that these two tear-film measures may not be agreeable [4, 6].

### 2.4.2. Fluorescein profilometry

Although TCR estimated with OCT-based measurements of tear meniscus height correlates with fluorescein tear clearance test and may have higher clinical utility than fluorophotometry, it does not provide the full insight into tear film dynamics. Fluorophotometric technique requires an expensive, specialized tools and sophisticated skills to perform measurements and as such it is mostly confined to research settings. An attempt was made to improve clinical utility of TTR assessment [96], where fluorescein profilometry-based technique of tear wash-out was introduced. This technique may be more easily applied in clinical environment, as it is less time-consuming, easier to perform, utilizes low concentration of fluorescent dye and, since the instrument can be used also to assess the corneoscleral topography, it may be useful in clinical environment as a multi-tool for topography, contact lens fitting and TCR estimations.

Fluorescein profilometry can be used to analyze subtle changes occurring in the tear film not only on a limited portion of the tear film like in fluorophotometry, but also on the exposed proportion of the anterior eye surface and with its high spatial resolution and relatively short period of exposure (up to one minute), it is not limited by corneal fluorescence. The analysis is based on the observation of the fringes projected on the whole eye surface with blue LED diodes. The main limitation of this approach is that tear clearance followed with this method cannot be considered quantitative, since fluorescein is instilled with the strip and thus the volume and concentration of substance instilled cannot be precisely determined. This issue needs to be further investigated and specified in the future studies. Nevertheless, it was not the first attempt to measure TCR after moistened fluorescein-impregnated strip was used to apply fluorescein [31]. Mean tear turnover rate (or so-called Fluorescein Wash-out rate) assessed in this study is displayed in the table 2. Those results seems to be similar to OCT based Tear Clearance Rate of young, healthy subjects and is also assumed as a manifestation of an early-phase tear turnover.

Table 2. Different tests for tear clearance assessment reported in literature measured by other means than fluorophotometry and scintigraphy

| Year | Report | Type (Number of subjects) | Method | Tear Turnover Rate [%/min] |
|---|---|---|---|---|
| 2014 | Zheng et al. [93] | Young group (30) | OCT-based TMH | 35.2 ± 11.0 |
|  |  |  | OCT-based TMA | 28.1 ± 12.4 |
|  |  | Elderly group (30) | OCT-based TMH | 12.4 ± 7.3 |
|  |  |  | OCT-based TMA | 6.2 ± 9.1 |
| 2017 | Garaszczuk, Iskander [96] | Normal (40) | Fluorescein profilometry | 39 ± 23 |
| 2017 | Garaszczuk et al. [94] | Normal (50) | OCT-based TMH | 29 ± 13 |

TMH – Tear meniscus height, TMA – tear meniscus cross-section area.

### 3. SUMMARY

After decades, the topic of tear turnover assessment is still a subject of investigation [107]. Tear turnover is described as a global measure of the integrity of the lacrimal system and tear exchange on the ocular surface and TTR, a temporal measure of tear turnover is shown to have a potential in dry eye disease diagnosis. TTR is also being considered a marker of ocular inflammation and was shown to be proportional to the sum of the effects of different processes occurring in the tear film and ocular tissues. Fluorophotometry-based evaluation of TTR, its limitations and TTR's impact on the ocular surface health were widely studied, however rarely performed in the clinical settings. Traditional, simpler clearance tests are either invasive or mostly based on subjective assessment and cannot follow dynamic changes naturally occurring in the tear film.

#### 3.1. Expert Commentary

Tear turnover is described as a global measure of the integrity of the lacrimal functional unit and tear exchange on the ocular surface. It is proportional to the sum of the effects of all the dynamic processes occurring in the tear film and an indirect measure of dry eye associated ocular surface irritation regardless of normal or reduced aqueous tear production. Tear turnover rate can be used for the differential diagnosis of dry eye disease. It was shown to correlate with multiple factors associated with dry eye and ocular surface inflammation. Values reported for dry eye subjects are two to five times lower than values reported for controls. Large variability of TRR was reported amongst subjects. Tear turnover was found to increase suddenly due to a variety of stimuli, like coughing, sneezing, wind or even psychological factors, like the effort to keep the eyes open, as well as variation in blinking rate in some individuals. The main weakness of standardized tear turnover assessment is that it is mostly confined to research settings, due to the complexity of the applied methodology and the need for specialized equipment. Standard assessment procedures for tear film diagnosis poorly correlate with subject reported symptom and there is still a need for developing new objective, quantitative methods to diagnose dry eye disease. The fluorescein clearance tests, developed to be a clinically applicable alternative for fluorophotometric measurements of tear turnover are less quantitative in nature and are often based on subjective evaluation or colorimetric comparison. Most of them do not allow following the complex tear film dynamics. Further research and development should provide ways to measure tear turnover in a more clinical, simpler, however still quantitative, reproducible and repeatable manner. New experimental techniques were proposed to simplify and facilitate the clinical application of tear turnover and tear clearance analysis. However, further investigation is needed to standardize these procedures and to allow quantitative, automatic measurements and image processing. The key issues are to make the methodology simple and automatic, without decreasing its ability to follow dynamic changes occurring in the tear film. An important factor that should also be taken into consideration is the inter-subject variability of tear volume prior to instillation. The assumed initial volume of 7 µl cannot be applied to all subjects. Also care should be taken not to induce forced blinking as it affects the characteristics of tear drainage. Reported large variations in blinking rate even amongst healthy subjects should also be considered. The initial volume of tears can be estimated by back-extrapolating to time zero the initial fluorescence decay assessed with fluorophotometry.

Development of new image-processing software, standardizing the amount of fluid to be instilled and automatic image-processing algorithms should facilitate the clinical application of tear turnover and tear clearance rate assessment.

4. Five-year view

Over the next 5 years, research should focus on developing standardized methodology to assess TTR as a clinical marker of ocular disease. This can be achieved by proposing simpler alternatives, using multipurpose tools that could be utilized to measure many different tear film characteristics or ocular parameters. Developing new image-processing software or enhancing image contrast are few means by which clinical utilization of tear turnover assessment could be realized. Also, standardizing the quantity and concentration of applied solutions to decrease their impact on tear film properties, minimizing the effect of reflex tearing, investigating the intra-day and visit-to-visit changes in tear flow are additional factor that should improve the current state of methods for assessing tear turnover. Future studies should focus on continuous measurement to allow following dynamic changes in the tear film and tear meniscus. Correlations of OCT-based TCR with other tear film measures and fluorophotometric technique can lead to better understanding of the complexity of tear film dynamics and the role of tear clearance in the pathogenesis of dry eye syndrome and ocular irritation. Future development may lead to using TTR as a standard test in dry eye and ocular surface disease diagnosis.

5. Key issues

- Tear turnover is described as a global measure of the integrity of the lacrimal functional unit and tear exchange on the ocular surface.
- Tear turnover rate (TTR), a temporal measure of tear turnover is proportional to the sum of the effects of tear secretion by the glands, fluid transudation through the conjunctiva, tear drainage through nasolacrimal duct, evaporation and conjunctival and corneal permeability to substances.
- TTR can be estimated based on direct or indirect observation of the decay of the dye placed in the tear fluid over time with the production and elimination of tears. Three main types of techniques used to assess TTR include:

    (1) Fluorescein tear clearance tests,

    (2) Techniques utilizing electromagnetic spectrum and tracer molecule,

    (3) The novel experimental techniques include utilizing optical coherence tomography and fluorescein profilometry.
- TTR was shown to be an indirect measure of dry eye associated ocular surface irritation, severity of corneal epithelial disease and was shown to be reduced in symptomatic dry eye subjects. Low TTR is also associated with aging, MGD, decreased ocular surface sensitivity and promotes ocular surface inflammation.
- The topic of TTR assessment was recently reintroduced and new techniques have been developed to propose less invasive, less time consuming and simpler methodologies for the assessment of tear dynamics that have the potential to be utilized in clinical practice for dry eye and ocular surface disease diagnosis.